\documentstyle[aps,twocolumn,eqsecnum,epsf,tighten]{revtex}

\newcommand{\be}{\begin{eqnarray}}  
\newcommand{\ee}{\end{eqnarray}}

\newcommand{\jtlea}[1]{\label{#1}}
\newcommand{\jtleq}[1]{\label{#1}}

\newcommand{\jtref}[1]{\ref{#1}} 
\newcommand{\jtcite}[1]{{\small{${\cite {#1}}$}}} 

\newcommand{\jtbibitem}[1]{\bibitem{#1}} 

\newcommand{\vect}[1]{\bf #1}

\newcommand{\cali}[1]{\cal #1}

\font\upright=cmu10 scaled\magstep1

\newcommand{\1}{\hbox{\upright\rlap{I}\kern -2.5pt 1}}
\def\square{\vcenter{\vbox{\hrule height.4pt
          \hbox{\vrule width.4pt height4pt
          \kern8pt\vrule width.4pt}\hrule height.4pt}}}

\begin{document}

\twocolumn[\hsize\textwidth\columnwidth\hsize\csname
@twocolumnfalse\endcsname

\typeout{--- Title page start ---}

\renewcommand{\thefootnote}{\fnsymbol{footnote}}


\begin{flushleft}\hspace{13.5cm}
DAMTP-1998-154 \\ \hspace{13.5cm}
hep-th/9811236 \\ \hspace{13.5cm}
~\\
\end{flushleft}

\title{Pressure of the Non-equilibrium $O(N)\; \Phi^{4}$ Theory
in the Large $N$ Limit }

\author{P.Jizba and  E.S.Tututi}

\address {DAMTP, University of Cambridge,
Silver Street, Cambridge, CB3 9EW, UK}

\date{\today}

\maketitle

\begin{abstract}\noindent 
We calculate the off-equilibrium hydrostatic  pressure for the $O(N)\;
\Phi^{4}$ theory to  the leading order  in $1/N$.   The present paper,
the first of a series, concentrates on  the calculation of pressure in
the    non-equilibrium  but  translationally  invariant   medium.  The
Jaynes-Gibbs  principle  of maximal entropy  is  used to introduce the
relevant  density  matrix  which  is  then directly   implemented into
dynamical   equations   through generalised Kubo-Martin-Schwinger (KMS)
conditions. We show that in the large $N$ limit use of Ward identities
enables the pressure  to  be expressed in  terms of  two point Green's
functions.   These  satisfy the  Kadanoff-Baym   equations  which are
exactly  solvable, and we  explicitly calculate the pressure for three
illustrative choices of $\rho$. 
\end{abstract}

\draft 
\vskip2.5pc]



\renewcommand{\thefootnote}{\fnsymbol{footnote}}
\setcounter{footnote}{0}


\typeout{--- Main Text Start ---}

\section{Introduction}
\label{sec1}

\noindent In  recent   years  significant progress  has been    made  in
understanding  the  behaviour   of    QFT  systems away   from   thermal
equilibrium. Motivation for the study  of such  systems comes both  from
the early  universe as well as  from  the quark-gluon plasma (deconfined
phase  quarks and gluons). Non-equilibrium   effects are expected to  be
relevant in the relativistic heavy-ion collisions planed at RHIC and LHC
in the near future \jtcite{TDL,SNW,HH,Hag}.

One  of    the   significant physical variables,   in     the context of
non-equilibrium  QFT, is  pressure.   Pressure, as  an easily measurable
parameter\footnote{In  this    connection  we may    mention  the piezo
resistive  silicon pressure sensors   used, for instance,  in superfluid
helium\jtcite{TH1,TH2} or  neutron (X-ray) diffraction technique used in
solid state physics\jtcite{SH,JSO}.}, is  expected to play  an important
role in a detection  of phase transitions.   This is usually ascribed to
the fact that the   pressure  should  exhibit  a discontinuity   in  its
derivative(s) when  the      local  phase  transition   occurs.      The
aforementioned has found its vindication in solid state physics and
fluid mechanics, and may  play a crucial role,  for instance, in various
baryogenesis scenarios. 

It is well  known that for systems  in thermal equilibrium, the pressure
may be   calculated   via the   partition  function  \jtcite{LW,LB1,ID}.
However,  this procedure cannot   be extended to off-equilibrium systems
where is no such thing as the grand-canonical potential.  In this letter
we consider   an  alternative definition  of    pressure,  based on  the
energy-momentum  tensor.     This, so  called,  hydrostatic  pressure is
defined as  the     space-like  trace of  the  energy-momentum    tensor
\jtcite{GMW,PJ},  and  in   equilibrium,  it  is   identified  with  the
thermodynamical pressure via the  virial theorem \jtcite{LW}.  There are
several problems with the validity of this identification on the quantum
level, indeed gauge theories suffer from a conformal (trace) anomaly and
require special   care\jtcite{LW}.    However,   we will   avoid    such
difficulties  by  focusing on  a  scalar theory  which  is  free of  the
mentioned complications \jtcite{LW,PJ}.  The major advantage of defining
pressure through the energy-momentum tensor stems from the fact that one
may effortlessly    extend the hydrostatic   pressure to non-equilibrium
systems (for discussion see Ref.\jtcite{PJ}). 

The aim of this paper  is to provide  a systematic prescription for  the
calculation of the hydrostatic pressure in  non-equilibrium media.  This
requires three concepts; the  Jaynes-Gibbs principle of maximal entropy,
the Dyson-Schwinger equations and the hydrostatic pressure.  In order to
keep  our discussion  as  simple as possible   we  illustrate the  whole
procedure on a  toy model system,  namely the  $O(N)\; \Phi^{4}$ theory.
The  latter  has advantage of  being exactly  solvable in the  large $N$
limit, provided that we deal with a translationally invariant medium. As
a result the hydrostatic pressure may be expressed in a closed form. 

In order  to provide meaningful  results  also for readers  not entirely
familiar with  the  Dyson-Schwinger   equations  and  the   Jaynes-Gibbs
principle of maximal entropy,  we briefly summarise  in Sec.II the basic
essentials.    (More    detailed    discussion   may     be  found    in
Refs.\jtcite{CH,Jayn2,J-T}.)  As     a   byproduct we     construct  the
generalised KMS  conditions. Sec.III   is devoted to   the study  of the
(canonical) energy momentum tensor in the $O(N)\;  \Phi^{4}$ theory.  If
both the density matrix  and the  full  Hamiltonian are invariant  under
$O(N)$ symmetry one obtains Ward's identities  in a similar manner as in
equilibrium.  We show how  these drastically simplify the expression for
the pressure.   In Sec.IV we concentrate our  analysis  on the large $N$
limit.  In  this setting we  derive  a  very  simple expression  for the
pressure -  pressure depends only on  two-point Green  functions.  Sec.V
then  forms the vital  part of this paper.   Owing to  the fact that the
infinite hierarchy of the Dyson-Schwinger equations is closed (basically
by chance) we  obtain simple equations for  two-point Green  functions -
Kadanoff-Baym equations.       These are   solved   exactly    for three
illustrative     density   matrices  $\rho$.    We  choose  deliberately
translationally invariant $\rho$'s.    The reason is twofold.   Firstly,
for  a  non-translationally invariant medium one   must use the improved
energy momentum tensor  instead of the canonical  one \jtcite{PJ}.  This
is rather  involved  and  it  will  be subject   of our   future  paper.
Secondly, the    Kadanoff-Baym  equations turn   out    to be hyperbolic
equations whose fundamental solution is well known.   As a result we may
evaluate, for  the  density matrices  at hand, the  hydrostatic pressure
explicitly.  The paper ends with a discussion.

\section{Basic formalism}
\label{sec2}

\noindent  The key object of  our interest is the energy-momentum tensor
$\Theta_{\mu \nu}(x)$. A  typical contribution to $\Theta_{\mu  \nu}(x)$
can be written as

\begin{equation}
{\cali{D}}_{\mu_{1}}\Phi(x)       \;  {\cali{D}}_{\mu_{2}}\Phi(x) \ldots
{\cali{D}}_{\mu_{n}}\Phi(x)\;.  \jtleq{CO1} 
\end{equation}

\noindent Here $\Phi$     is a field in   the    Heisenberg picture  and
${\cali{D}}_{\mu_{i}}$ stands for a corresponding differential operator.
Since   ${\cali{D}}_{\mu_{i}}\Phi(x)$ and  ${\cali{D}}_{\mu_{j}}\Phi(x)$
generally do not commute for $i\not= j$, one must prescribe the ordering
in $\Theta_{\mu \nu}$. Our strategy is based on the observation that one
can conveniently define such ordering via the non-local operator 

\begin{eqnarray}
&&{\lim_   {x_i\to   x}T^{*}\{   {\cali{D}}_{\mu_{1}}\Phi  (x_{1})\cdots
{\cali{D}}_{\mu_{n}}\Phi     (x_{n})      \}           =}\nonumber    \\
&&\mbox{\hspace{1.3cm}} \lim_ {x_i\to x} {\cali{D}}(i\partial_{\{\mu\}})
T\{\Phi (x_{1})\cdots \Phi (x_{n})\}, \jtleq{TP1} 
\end{eqnarray}

\noindent  where   ${\cali{D}}(i\partial_{\{\mu\}})$  is  just  a useful
short-hand  notation for ${\cali{D}}_{\mu_{1}}{\cali{D}}_{\mu_{2}}\ldots
{\cali{D}}_{\mu_{n}}$, and $T^{*}$ is defined in such a way \jtcite{CCR}
that for the product of fields without derivatives  it is simply the $T$
product, whilst for fields containing  derivatives it  is the usual  $T$
product with all the derivatives pulled out  of the $T$-ordering symbol.
It  is  clear that both  $T^{*}$ and  $T$ coincide  if all the arguments
$x_{i}$ are  different, so  $(T^{*}-T)(\ldots)$  is an  operator with  a
support at the contact points. The $T^{*}$ ordering is in general a very
useful tool whenever one deals  with composite operators.  In the sequel
we shall repeatedly use this fact.

\subsection{Off-equilibrium Dyson-Schwinger equations}

\noindent  Let   us   now briefly   outline   the   derivation  of   the
Dyson-Schwinger equations  for  systems  away  from  equilibrium.    For
simplicity we illustrate this on a single scalar field $\Phi$. 

We start with the action  $S[\Phi]$ where $\Phi$  is linearly coupled to
an external  source $J(x)$.  Working with the  fields  in the Heisenberg
picture, the operator equation of motion can be written as 

\begin{equation}
\frac{\delta S}{\delta \Phi(x)}[\Phi = \Phi^{J}] + J(x) = 0, \jtleq{4.0} 
\end{equation}

\noindent where  the  index $J$ emphasises    that $\Phi$ is  implicitly
$J$-dependent.   This dependence  will be  made  explicit  via a unitary
transformation connecting $\Phi^{J}$  (governed  by  $H - J\Phi$)   with
$\Phi$ (governed by $H$). If $J(x)$ is switched  on at time $t=t_{i}$ we
have   $\Phi^{J}(x)=T_{C} \{\Phi(x)\; \mbox{exp}\;(i\int_{C}d^{4}y    \;
J(y)\Phi(y))\}$, where the  close-time path $C$ - the Keldysch-Schwinger
path - runs  along  the real axis  from   $t_i$ to $t_{f}$  ($t_{f}$  is
arbitrary, $t_{f}>t_{i}$) and then back  to $t_i$. With this setting  we
can rewrite (\jtref{4.0}) as



\begin{eqnarray}
&&T^{*}_{C}\left(   \left\{ \frac{\delta  S[\Phi_{\pm}]}{\delta\Phi}   +
J_{\pm}\right\}\right.   \nonumber\\ &&\left.  \times \exp(i \int d^{4}y
\; (J_{+}(y)\Phi_{+}(y) - J_{-}(y)\Phi_{-}(y))) \right) = 0\, , 
\end{eqnarray}

\noindent where,  as usual \jtcite{LW,KCC}, we  have labelled  the field
(source) with the time argument on the upper branch of $C$ as $\Phi_{+}$
($J_{+}$) and that with the time argument on the  lower branch of $C$ as
$\Phi_{-}$ ($J_{-}$).    Introducing   the metric  $(\sigma_{3})_{\alpha
\beta}$ ($\sigma_{3}$ is the   usual Pauli matrix  and $\alpha,  \beta =
\{+;-\}$)     we  can    write   $J_{+}\Phi_{+}     -  J_{-}\Phi_{-}   =
J_{\alpha}\;(\sigma_{3})^{\alpha         \beta}\;    \Phi_{\beta}      =
J^{\alpha}\;(\sigma_{3})_{\alpha \beta}\; \Phi^{\beta}$. For the  raised
and lowered indices we simply read:  $\Phi_{+} = \Phi^{+}$ and $\Phi_{-}
= -\Phi^{-}$ (similarly for $J_{\alpha}$). Taking $\mbox{Tr}(\rho \ldots
)$ with  $\rho  = \rho[\Phi, \partial  \Phi,  \ldots]$ being the density
matrix, we obtain 

\begin{eqnarray}
\frac{1}{Z[J]}\frac{\delta S}{\delta \Phi(x)} \left[\Phi^{\alpha}(x) = -
i\frac{\delta}{\delta J_{\alpha}(x)}\right]Z[J]   = -J^{\alpha}(x)  \, ,
\jtlea{4.15} 
\end{eqnarray}

\noindent with $Z[J]=  {\rm Tr}\left\{ \rho  \; T_{C} \mbox{exp}\;\left(
i\int_{C}d^{4}y  \;  J(y)\Phi(y) \right)\right\}$  being the  generating
functional of Green's   functions.  Employing the commutation  relation:
$-i\frac{\delta   }{\delta  J_{\alpha}}Z   =  Z\,(  \phi^{\alpha}   -  i
\frac{\delta}{\delta J_{\alpha}})$, we may recast (\ref{4.15}) into more
convenient form, namely 

\begin{eqnarray}
-J^{\alpha}(x)   &=&   \frac{\delta         S}{\delta     \Phi(x)}\bigg[
\phi^{\alpha}(x)      \nonumber\\        &+&      i    \int     d^{4}y\;
G_{\beta}^{\alpha}(x,y)\;          (\sigma_{3})^{\beta           \gamma}
\;\frac{\delta}{\delta\phi^{\gamma} (y)} \bigg]\,\,\1 \, , \jtlea{4.16} 
\end{eqnarray}

\noindent where the  symbol $\1$ indicates  the unit. As usual, the mean
field, $\phi_{\alpha}$, is defined as the expectation value of the field
operator:    $\phi_{\alpha}(x)\equiv\langle   \Phi_{\alpha}(x) \rangle$.
Defining $Z[J] = \mbox{exp}(iW[J])$,  two-point Green functions are then
defined     as           $G_{\alpha\beta}(x,y)=      -\delta^{2}W/\delta
J^{\alpha}(x)\delta          J^{\beta}(y)$.     Setting    $J$        in
(\jtref{4.15})-(\jtref{4.16}) to $0$ (i.e. physical condition) we obtain
a first out of  infinite hierarchy of  equations for  Green's functions.
Successive   $J$ variations  of  (\jtref{4.15})-(\jtref{4.16})  generate
higher order equations in the hierarchy.  The  system of these equations
is usually referred to as the Dyson-Schwinger equations.

For the future  reference is convenient to have  the expression  for the
effective action $\Gamma [\phi_{c}]$.  This is connected with $W[J]$ via
the Legendre transform: 

\begin{equation}
\Gamma[\phi_{c}] =  W[J]  -  \int_{C}  d^{4}y  \; J(y)   \phi_{c}(y)\, .
\jtlea{letrans} 
\end{equation}

\noindent Following  the  previous reasonings,  one can  easily persuade
oneself that the expectation value of $\Theta^{\mu \nu}$ reads 

\begin{eqnarray}
\langle     \Theta^{\mu  \nu}(x)   \rangle     &=&  \langle  \Theta^{\mu
\nu}[\Phi(x)] \rangle = \Theta^{\mu \nu}[\phi_{  +}(x) \nonumber\\ &+& i
\int           d^{4}y\;         G_{+    \beta}(x,y)\;(\sigma_{3})^{\beta
\gamma}\;\frac{\delta}{\delta  \phi^{\gamma}   (y)}]\,\,   \1    \,    .
\jtleq{exvaten} 
\end{eqnarray}

\noindent  We have automatically used  the sub-index `$+$' as the fields
involved in $\Theta^{\mu \nu}$ have, by definition, the time argument on
the upper branch of $C$.


\subsection{The Jaynes-Gibbs principle of maximal entropy}


\noindent  Another important  concept  to be  used   is the Jaynes-Gibbs
principle of maximal entropy. This  prescription allows one to construct
the density matrix in such a way that the informative content of $\rho$,
subject to certain constraints imposed by  our theory/experiment, is the
least \jtcite{Jayn2,J-T,d}.  The  practical side of the principle  rests
on the  maximalisation of the information  (or Shannon) entropy, defined
as $S[\rho] = - \mbox{Tr}(\rho \;\mbox{ln}\rho)$, subject to constraints
imposed  by  our knowledge  of expectation  values of  certain operators
$P_{i}[\Phi,  \partial \Phi]$.  In  contrast to thermal equilibrium, the
$P_{i}[\ldots]$'s need  not to be  the  constants of  motion (space-time
dependences are allowed). 

In practice the Jaynes-Gibbs   principle is used  in the  following way;
given the constraints 

\begin{equation}
\langle  P_{k}[\Phi,     \partial    \Phi](x_{1},  \ldots)    \rangle  =
g_{k}(x_{1},\ldots),\;\;\;\;\ k= 1\ldots n\;, \jtlea{con1} 
\end{equation}

\noindent   the maximum of the   information entropy then determines the
most plausible $\rho$ as

\begin{equation}
\rho                 =               \frac{1}{{\cali{Z}}[\lambda_{k}]}\;
\mbox{exp}(-\sum_{k=1}^{n}\prod_{j}d^{4}x_{j}\;       \lambda_{k}(x_{1},
\ldots)P_{k}[\Phi,\partial \Phi]), \jtlea{rho1} 
\end{equation}

\noindent where the `partition function' ${\cali{Z}}$ is 

\begin{displaymath}
{\cali{Z}}[\lambda_{k}]                                                =
\mbox{Tr}\left\{\mbox{exp}(-\sum_{k=1}^{n}\prod_{j}d^{4}x_{j}\;
\lambda_{k}(x_{1}, \ldots)P_{k}[\Phi,\partial \Phi])\right\}. 
\end{displaymath}

\noindent  In both previous expressions the   time integration is either
not present at all  (so $f_{k}$ are specified only  at the initial  time
$t_{i}$),  or is taken over  the gathering interval ($-\tau, t_{i}$). If
instead,  one   has  knowledge     about  the  expectation  values    of
$P_{i}[\ldots]$ at some   discrete  times  $t_{l}$, the    corresponding
integration must be replaced by a discrete summation. In the same manner
as  at   equilibrium  one may     eliminate  the  Lagrange   multipliers
$\lambda_{k}$ by solving the $n$ simultaneous equations 

\begin{equation}
g_{k}(x_{1},  \ldots) =   - \frac{\delta \;  \mbox{ln}{\cali{Z}}}{\delta
\lambda_{k}(x_{1}, \ldots )}.  \jtlea{jg3} 
\end{equation}

\noindent The solutions of (\jtref{jg3}) may be formally written as

\begin{equation}
\lambda_{k}(x_{1},\ldots )   = \frac{\delta  \;  S_{G}[g_{1},   \ldots ,
g_{n}]|{{}_{max}}}{\delta g_{k}(x_{1}, \ldots)}.  \jtlea{ent1} 
\end{equation}

\noindent Now, in order to reflect  the density matrix (\jtref{rho1}) in
the  Dyson-Schwinger equations, we   need to construct the corresponding
boundary conditions.  This   may be done quite straightforwardly.  Using
the cyclicity of $\mbox{Tr}(\ldots)$ together with the relation

\begin{displaymath}
e^{A}Be^{-A}       =       \sum_{n=0}^{\infty}\frac{1}{n!}C_{n},\;\;\;\;
 C_{0}=B,C_{n}= [A,C_{n-1}], 
\end{displaymath}

\noindent we can write the  generalised KMS conditions for the $n$-point
Green function as 

\begin{eqnarray}
&&\langle   \Phi(x_{1})\ldots \Phi(x_{n}) \rangle  = \langle \Phi(x_{2})
\ldots        \Phi(x_{n})\Phi(x_{1})\rangle             \nonumber     \\
&&\mbox{\hspace{1.55cm}}+     \sum_{k=1}^{\infty}\frac{1}{    k!}\langle
\Phi(x_{2})\ldots \Phi(x_{n})C_{k}(x_{1}) \rangle \,, \jtleq{CMS11} 
\end{eqnarray}

\noindent  where  $A=  {\mbox{ln}}(\rho)$, $B   = \Phi(x_{1})$, $x_{10}=
t_{i}$.   So   for the two-point  Green   functions  we have\footnote{In
special cases when  $\rho = |0\rangle \langle  0|$ or  $\rho = e^{-\beta
H}/{\cali{Z}}(\beta)$ the boundary conditions are the well known Feynman
and KMS boundary conditions, respectively.} 

\begin{displaymath}
G_{+-}(x,y)                    =          G_{-+}(x,y)                  +
\sum_{n=1}^{\infty}\frac{1}{n!}\mbox{Tr}(\rho \;\phi(x)C_{n}(y)). 
\end{displaymath}

\noindent In this paper  and the paper to  follow we aim to  demonstrate
that conditions  (\jtref{CMS11})  together with  the causality condition
are sufficient to determine Green's functions uniquely.

\section{The $O(N)\; \Phi^{4}$ theory}
\label{sec3}


\noindent  The  $O(N)\;  \Phi^{4}$ theory   is   described by   the bare
Lagrangian

\begin{equation}
{\cali{L}}=         \frac{1}{2}\sum_{a=1}^{N}\left(            (\partial
\Phi^{a})^{2}-m_{0}^{2}(\Phi^{a})^{2}              \right)             -
\frac{\lambda_{0}}{8N}\left(  \sum_{a=1}^{N} (\Phi^{a})^{2} \right)^{2}.
\jtleq{6} 
\end{equation}

\noindent   The   corresponding   canonical   energy-momentum  tensor is
$\Theta^{\mu \nu}  =\sum_{a}\partial^{\mu}\Phi^{a}\partial^{\nu}\Phi^{a}
- g^{\mu  \nu}{\cali{L}}$, and from  Eqs.(\ref{TP1}) and (\ref{exvaten})
its expectation value is 

\begin{eqnarray}
&&\Theta^{\mu\nu}(\Phi_{+}(x))=\langle \Theta^{\mu \nu}_{C}(x) \rangle =
i\sum_{c}\partial_{x}^{\mu}\partial_{y}^{\nu}\;G^{cc}_{++}(x,y)|_{x=y}
\nonumber\\
&&\mbox{\hspace{7mm}}-\frac{g^{\mu\nu}i}{2}\sum_{c}\bigg\{\partial_{x}^{\alpha}\partial_{\alpha
y}\;G_{++}^{cc}(x,y)|_{x=y}        \nonumber\\    &&\mbox{\hspace{7mm}}-
m_{0}^{2}G_{++}^{cc}(x,x)\bigg\}+    \frac{g^{\mu   \nu}\lambda_{0}}{8N}
\sum_{c,d}        \bigg   \{\bigg((\phi^{c}_{+}(x))^{2}      \nonumber\\
&&\mbox{\hspace{7mm}}+                     iG^{cc}_{++}(x,x)\bigg)\bigg(
(\phi^{d}_{+}(x))^{2}       +      iG^{dd}_{++}(x,x)\bigg)   \nonumber\\
&&\mbox{\hspace{7mm}}+ 2    \bigg(    \phi^{c}_{+}(x)\phi^{d}_{+}(x)   +
iG^{cd}_{++}(x,x)\bigg)iG^{cd}_{++}(x,x)\bigg\}              \nonumber\\
&&\mbox{\hspace{7mm}}+  \mbox{terms  with $\Gamma^{3}$ and $\Gamma^{4}$}
\jtleq{valten} \, . 
\end{eqnarray}

\noindent  Before  proceeding further  with  our development, we want to
show how  one  can  significantly simplify  Eq.(\jtref{valten}) provided
that both the   density matrix and the  Hamiltonian  are invariant under
rotations in the $N$-dimensional vector  space of fields. This situation
would occur if the system was  initially prepared in such  a way that no
field $\Phi^{a}$  was  favoured over another.  The  fact  that $\rho$ is
invariant under $O(N)$ transformations means that 

\begin{equation}
U(\epsilon)\rho[\Phi,      \partial     \Phi,  \ldots]U^{-1}(\epsilon) =
\rho[\Phi, \partial \Phi, \ldots]\;, \jtlea{e33} 
\end{equation}

\noindent where  the  fields $\Phi^{a}$  transform under $N$-dimensional
rotations: $U(\epsilon)\Phi^{a}U^{-1}(\epsilon)   =  R^{-1}(\epsilon)^{a
b}\Phi^{b}     =[\mbox{exp}(\epsilon_{i}T_{i})]^{a b}\Phi^{b}$,    where
$R(\epsilon)$ is the   rotation  matrix in the  $N$-dimensional   vector
space,  and the generators $T_{i}$  are real and antisymmetric $N \times
N$ matrices.   It is obvious  that the  previous relation for $\Phi^{c}$
can be satisfied  for all  times only  if  the full  Hamiltonian,  which
governs the evolution  of   $\Phi^{a}$, is  also  invariant against  the
$N$-dimensional rotations. 

Let us now  consider the generating  functional $Z[J]$ corresponding  to
the $O(N)$  symmetric density matrix.  Employing  the cyclic property of
$\mbox{Tr}(\ldots)$  together  with the    infinitesimal transformation,
$\delta     R(\epsilon)= 1 +    \epsilon_{i}T_{i}$,  we  obtain that the
variation   of $Z$   must  vanish.   The  latter  implies  the following
(unrenormalised) Ward's identities:

\begin{eqnarray}
\int_{C} d^{4}y \;J^{a}(y)    \frac{\delta   W[J]}{\delta  J^{b}(y)}   =
\int_{C}  d^{4}y\;   J^{b}(y) \frac{\delta  W[J]}{\delta  J^{a}(y)}\,  .
\jtlea{e7} 
\end{eqnarray}

\noindent Taking successive variations with   respect to source $J$,  we
obtain  the following results   (see also \jtcite{J-T}): $n$-point Green
functions with $n$ odd vanish, while for $n$ even ($n=2k, k=1,2,\ldots$)
one has 

\begin{eqnarray}
&&G^{a_1a_2\dots        a_{2k}}_{\alpha_1\cdots\alpha_{(2k)}}(x_1,\ldots
,x_{2k}) \nonumber\\   &&\qquad\qquad = \sum_{{\rm  all\,\, pairings}}\,
\prod_{i<j}                                              \delta_{a_ia_j}
G^{(2k)}_{\alpha_1\cdots\alpha_{2k}}(x_1,\ldots      ,x_{2k})   \,     ,
\jtleq{apc8} 
\end{eqnarray}

\noindent where $G^{(2k)}$ is   a universal $2k$-point  Green  function.
Similar      results     \jtcite{J-T}    can       be      obtained  for
$\Gamma^{a_{1}a_{2}\ldots            a_{2k}}_{\alpha_{1}\alpha_{2}\ldots
\alpha_{2k}}(\ldots)$.

Finally  note    that  these results  enable     the expression  for the
expectation   value   of the energy   momentum   tensor to be simplified
significantly to

\begin{eqnarray}
\langle    \Theta^{\mu\nu}(x)\rangle      &=&         iN\partial^{\mu}_x
\partial^{\nu}_y\,\,  G_{++}(x,y)\big\vert_{x=y}   \nonumber\\  &-& {N+2
\over  8}\lambda_0 g^{\mu\nu}(G_{++}(x,x))^2  \nonumber\\ &-&   i{N\over
2}g^{\mu\nu}\bigg( \partial^{\mu}_x \partial^{\nu}_y\,\, G_{++}(x,y)\big
\vert_{x=y} \nonumber\\   &-&   m_0^2  \,\,G_{++}(x,x) \bigg)   +   {\rm
terms\,\,\, with}\,\,\,\, \Gamma^{(4)}\, .  \jtlea{me} 
\end{eqnarray}

\noindent In the rest  of this paper  we shall confine ourselves only to
situations where both $\rho$ and $H$ are $O(N)$ invariant.

\section{The large $N$ limit}
\label{sec4}

\noindent  Let us now examine  behaviour  of  (\jtref{me}) to the  order
${\cali{O}}(1/N)$.  For this purpose it is important  to know how either
$G^{(n)}$ or $\Gamma^{(n)}$ behave  in the $N \rightarrow \infty$ limit.
At  $T=0$ or in  equilibrium     the  Feynman diagrams are     available
\jtcite{ID,PJ}  and the corresponding combinatorics   can be worked  out
quite simply.  On the other hand, the situation in off-equilibrium cases
is more difficult as we do not have at our disposal Wick's theorem.  One
may  devise    various  diagrammatic   approaches, e.g.    kernel method
\jtcite{CH},    cumulant  expansion  \jtcite{M},    correlation diagrams
\jtcite{PH}, etc.  Instead of relying on any graphical representation we
show   that for  both  equilibrium   and   off-equilibrium systems,  the
situation may be approached from far more  general standpoint using only
Ward's identities and properties of $\Gamma$ and $W$.

In order to find the leading behaviour at large $N$ it is presumably the
easiest to consider    the Legendre transform  (\jtref{letrans}).    The
explicit $N$ dependence  may be obtained by  setting $\phi^{c} =  \phi$,
which implies $J^{c}=J$  for all the group indices. Eq.(\jtref{letrans})
then indicates that both $\Gamma[\phi]$ and $W[J]$ are of order $N$.  If
we expand $\Gamma[\phi^{a}]$ in terms  of $\phi^{c}$ taking into account
Ward's identities we get

\begin{eqnarray}
\Gamma[\phi]  = \Gamma[0] &+& \frac{1}{2}  N \int_{C} d^{4}x\; d^{4}y \;
\Gamma^{(2)}(x,y) \; \phi(x)\phi(y) \nonumber  \\ &+& \frac{3}{4!} N^{2}
\int_{C} d^{4}x\;  d^{4}y\;  d^{4}z\;  d^{4}q  \;  \Gamma^{(4)}(x,y,z,q)
\nonumber\\ &\times  & \phi(x)  \phi(y)  \phi(z) \phi(q) +  \cdots  \, .
\jtlea{W2} 
\end{eqnarray}

\noindent Since the LHS of (\jtref{W2})  is of order $N$, $\Gamma^{(2)}$
must be  of order $N^{0}$,  $\Gamma^{(4)}$  of order  $N^{-1}$, and,  in
general, $\Gamma^{(2n)}$ must be of order $N^{1-n}$.  This suggests that
in the expression  for  the energy-momentum tensor (\jtref{me}),   terms
containing  $\Gamma^{(4)}$ can be  ignored.  The  above  argument can be
repeated in a similar way for $W$. 

Hence,  collecting our results together,   the expectation value of  the
energy-momentum tensor to leading order in $N$ may be written as

\begin{eqnarray}
&&\langle            \Theta^{\mu\nu}(x)\rangle\nonumber               \\
&&\nonumber{\hspace{5mm}}=    iN\partial^{\mu}_x    \partial^{\nu}_y\,\,
G_{++}(x,y)\big\vert_{x=y}       -  {N      \over            8}\lambda_0
g^{\mu\nu}(G_{++}(x,x))^2 \nonumber    \\     &&\nonumber{\hspace{5mm}}-
i{N\over   2}g^{\mu\nu}\bigg(     \partial^{\mu}_x  \partial^{\nu}_y\,\,
G_{++}(x,y)\big \vert_{x=y}- m_0^2 \,\,G_{++}(x,x) \bigg).\\ \jtleq{me2} 
\end{eqnarray}

\noindent This  result is surprisingly  simple: the expectation value of
the  energy-momentum   tensor, and  thus  the hydrostatic   pressure, is
expressed purely in terms  of  two-point Green's functions.  The  latter
can be  calculated through the Dyson-Schwinger equations (\jtref{4.16}).
Furthermore, these equations have a very  simple form provided that both
the large $N$ limit and Ward's identities are applied.   If we perform a
variation of (\jtref{4.16}) with respect to $J_{\beta}(y)$ we obtain, to
order ${\cali{O}}(1/N)$,   the following  Dyson-Schwinger  equations for
two-point Green functions:

\begin{eqnarray}
&&\left(  \Box   +  m_{0}^{2} +  \frac{i  \lambda_{0}}{2}   \; G_{\alpha
\alpha}(x,   x)   \right)    G_{\alpha  \beta}(x, y)    \nonumber\\   &&
\mbox{\vspace{2cm}}  =  -\delta(y-x)(\sigma_{3})_{\alpha \beta}   \,   .
\jtleq{em5} 
\end{eqnarray}

\noindent These dynamical equations  for  two-point Green functions  are
better known as the Kadanoff-Baym equations \jtcite{KB}. 

Let  us mention  one more   point.  The  generalised KMS conditions  for
$G_{\pm \mp}$ are significantly simple in  the large $N$ limit.  This is
because in sum  (\jtref{CMS11}) only terms of  order ${\cali{O}}(N^{0})$
contribute.   This  implies  that only quadratic  operators $P_{i}[\Phi,
\partial \Phi]$ in  the density matrix  are relevant.  As a  result, the
Jaynes-Gibbs  principle naturally provides  a vindication of the popular
Gaussian Ansatz \jtcite{EJY,M1,M2}.

\section{Out-of-Equilibrium Pressure}
\label{sec5}

\noindent The   objective of  the present section   is to  show how  the
outlined mathematical  machinery works in  the  case of the  hydrostatic
pressure.   In order   to gain   some  insight  we    start with  rather
pedagogical,  but  by  no     means trivial  examples;   translationally
invariant,   non-equilibrium density  matrices.   We  consider  the more
difficult  case of translationally  non-invariant density  matrices in a
future paper. 

\subsection{Equilibrium}

\noindent As an important special case we can choose the constraints 

\begin{equation}
\langle P_{k}[\Phi,    \partial \Phi]  \rangle    |_{t_{i}} =   g_{k}  =
\mbox{constant}, \jtlea{eq1} 
\end{equation}

\noindent where $t_{i}$ is arbitrary. Eq.(\jtref{eq1}) then implies that
$P_{k}$'s are integrals of motion.   Since in the finite volume  systems
the spatial translational invariance is destroyed,  the only integral of
motion  (apart  from conserved charges)  is  the  Hamiltonian.  Thus the
system  is   in  thermal equilibrium   and the   laws  of thermodynamics
\jtcite{M} prescribe that $g  = \int^{T}_{0} dT' C_{V}(T')$  ($C_{V}$ is
the heat capacity at  constant volume $V$  and $T$ is temperature).  Eq.
(\jtref{ent1}) determines the   Lagrange  multiplier; $\lambda =  1/T  =
\beta$.  The density matrix maximising  the  $S_{G}$ is then the density
matrix  of   the  canonical  ensemble: $\rho  =  \frac{\mbox{exp}(-\beta
H)}{{\cali{Z}[\beta]}}$.  Due to the translational invariance of $\rho$,
the Kadanoff-Baym equations read 

\begin{equation}
\left(    \Box_{x}  +    m_{r}^{2}(T)\right)   G_{\alpha   \beta}(x-y) =
-\delta(x-y)(\sigma_{3})_{\alpha \beta} \, , \jtlea{em7} 
\end{equation}

\noindent where the temperature-dependent renormalised mass is (see, for
example    \jtcite{ID,PJ});  $m_{r}^{2}(T)  =    m_{0}^{2}   +   \frac{i
\lambda_{0}}{2} \;    G_{\alpha  \alpha}(0)$.   The    corresponding KMS
boundary condition is

\begin{equation}
G_{+-}({\vect{x}},t_{i}; {\vect{y}},0) = G_{-+}({\vect{x}},t_{i}-i\beta;
{\vect{y}},0).  \jtlea{em8} 
\end{equation}

\noindent Because $m_{r}(T)$ is a  spatial constant, a Fourier transform
solves equations  (\jtref{em7}).  The solutions of (\jtref{em7}) subject
to   condition (\jtref{em8}) are   then the  resumed  propagators in the
Keldysh-Schwinger formalism 

\begin{eqnarray}
iG_{\pm \pm}(k) &=& \frac{\pm  i}{k^{2} - m^{2}_{r} \pm  i\varepsilon} +
2\pi  f(|k_{0}|)\delta(k^{2}-m^{2}_{r})\nonumber \\ iG_{\pm \mp}(k)  &=&
2\pi  \left\{  \theta(\mp k_{0})   + f(|k_{0}|)   \right\}\delta(k^{2} -
m^{2}_{r})\; , \jtlea{sol1} 
\end{eqnarray}
  
\noindent   with  $f(x)  =   (\mbox{exp}(\beta   x)-1)^{-1}$  being  the
Bose-Einstein distribution. 

\vspace{3mm} 

\noindent Now,  the   total hydrostatic pressure  in  $d$  dimensions is
classically defined as \jtcite{LW,GMW,PJ} 

\begin{displaymath}
{\cali{P}}(T) = -\frac{1}{(d-1)}\langle \Theta^{i}_{\; i} \rangle. 
\end{displaymath}

\noindent Because $\Theta^{\mu \nu}$  is a composite operator, a special
renormalisation is  required \jtcite{LW,PJ,collins,brown}.   As  we have
shown    in   \jtcite{PJ}, for   translationally  invariant  systems the
renormalised pressure coincides with the, so called, thermal interaction
pressure $P$.  The latter reads 

\begin{eqnarray}
P(T) &=& {\cali{P}}(T) -  {\cali{P}}(0)\nonumber \\ &=& -\frac{1}{(d-1)}
\left\{ \langle \Theta_{\; i}^{i} \rangle  - \langle 0| \Theta_{\;i}^{i}
| 0 \rangle \right\}.  \jtleq{ppp10} 
\end{eqnarray}

\noindent  Result (\jtref{ppp10}) deserves   two comments.  Firstly, the
energy momentum tensor  $\Theta^{\mu \nu}$ need  not to be the canonical
one (however, the canonical  one is usually the  simplest one).  This is
because  energy  momentum tensors  are  mutually  connected  via Pauli's
transformation, i.e.

\begin{eqnarray}
&&{\tilde{\Theta}}^{\mu    \nu}         =     \Theta^{\mu  \nu}        +
\partial_{\lambda}X^{\lambda \mu \nu}\nonumber \\ && X^{\lambda \mu \nu}
= -X^{\mu \lambda \nu}, \jtleq{pp1} 
\end{eqnarray}

\noindent and owing to the space-time translational invariance

\begin{equation}
\langle {\tilde{\Theta}}^{\mu  \nu}    \rangle  =  \langle   \Theta^{\mu
\nu}\rangle + \partial_{\lambda} \langle X^{\lambda  \mu \nu } \rangle =
\langle \Theta^{\mu \nu}\rangle.  \jtlea{ppp8} 
\end{equation}

\noindent (The analogous  identity is  naturally  true at $T=0$.)  As  a
second point we should mention that prescription (\jtref{ppp10}) retains
its validity  for non-equilibrium media  as well.   This is  because the
short-distance behaviour of $G_{+ +}(x,y)$, which is responsible for the
singular behaviour of  $\Theta^{\mu  \nu}$,  comes from  the  particular
solution of the Kadanoff-Baym equation (\jtref{em7}).  The latter can be
chosen to be completely independent of the boundary conditions (actually
it is useful to chose the Feynman causal solution).   On the other hand,
the homogeneous solution of  (\jtref{em7}),  which is regular at  $|x-y|
\rightarrow 0$, reflects all the boundary conditions.   One may see then
that  the  UV singularities which    trouble $\Theta^{\mu \nu}$  may  be
treated in the same way as at the $T=0$.  Incidentally, the former is an
extension of the  well known fact that  in order to renormalise a finite
temperature QFT, it suffices to renormalise it at $T=0$.

Inserting the   solution  (\jtref{sol1})  into the  expression   for the
energy-momentum   tensor   (\jtref{me2})   we   arrive  at   the thermal
interaction pressure per particle (see also \jtcite{PJ,ACP}) 

\begin{eqnarray}
P(T) &=& \frac{T^{4}\;  \pi^{2}}{90} - \frac{T^{2}\;m^{2}_{r}(T)}{24}  +
\frac{T\;m^{3}_{r}(T)}{12  \pi}\nonumber   \\   &&  \nonumber   \\   &+&
\frac{\lambda_{r}}{8}\left(            \frac{T^{4}}{144}               -
\frac{T^{3}\;m_{r}(T)}{24   \pi}   +       \frac{T^{2}\;m^{2}_{r}(T)}{16
\pi^{2}}\right)\nonumber     \\   &&\nonumber \\   &+&  {\cali{O}}\left(
m^{4}_{r}(T)\;    \mbox{ln}\left(  \frac{m_{r}(T)}{T4   \pi     }\right)
\right)\,, \jtleq{b9} 
\end{eqnarray}

\noindent where the renormalised  coupling constant  $\lambda_{r}$ comes
from  the  $T=0$  renormalisation prescription:   $\Gamma^{(4)}_{aa  \to
bb}(s=0)=-\lambda_{r} /N$ ($s$  is  the usual  Mandelstam variable).   A
direct   calculation  of   $\Gamma^{(4)}$ in     the  large  $N$   limit
\jtcite{ID,PJ}, gives the relation for the renormalised coupling: 

\begin{equation}
\lambda_{r}   =    \lambda_0  -{1\over 2}    \lambda_0\lambda_{r}{1\over
(4\pi)^{d/2}} \Gamma\left[2-d/2\right] \left(m_{r}^{2}\right)^{d/2-2} \,
.  \jtleq{re6} 
\end{equation}

\noindent To regularise the theory we have used the usual MS scheme with
the dimensional regularisation ($d  \not=   4$).  The high   temperature
expansion of the pressure  (\jtref{b9}) to all   orders can be  found in
\jtcite{PJ}  where  the calculations, however,  were  approached  from a
different standpoint.

\subsection{Off-equilibrium I}

\noindent  The next  question   to   be  addressed  is  how the    above
calculations are modified in the  non-equilibrium case.  To see that let
us choose the following constraint 

\begin{equation}
g({\bf  k})=\langle   {\cali{H}}({\bf  k})\rangle|_{t_{i}}    =  \langle
{\tilde{\cali{H}}}({\bf {k}}) \rangle|_{t_{i}}.  \jtlea{ppp9} 
\end{equation}

\noindent   Here  $t_{i}$ is  arbitrary,  and  function  $g({\bf k})$ is
specified by  theory/experiment.  The ${\tilde{\cali{H}}}({\bf {k}})$ is
a     quadratic  operator      fulfilling   the    condition    $\langle
{\cali{H}}\rangle|_{N       \rightarrow      \infty}       =     \langle
{\tilde{\cali{H}}}\rangle$.  As    $g({\bf{k}})$      is  finite,   both
${\cali{H}}$  and ${\tilde{\cali{H}}}$  must  be renormalised  (i.e.  we
must subtract the zero temperature part). Obviously

\begin{displaymath}
{\tilde{\cali{H}}}({\bf{k}})                                           =
\omega_{k}a^{\dagger}({\bf{k}})a({\bf{k}}), 
\end{displaymath}

\noindent with $\omega_{k} = \sqrt{{\bf{k}}^{2}+ {\cali{M}}^{2}}$ and

\begin{equation}
{\cali{M}}^2  =m^2_0+{i\over   2}\lambda_0    \int {d^{d}k^{\prime}\over
(2\pi)^{d}} G_{++}(k^{\prime}) \, .  \jtleq{re3} 
\end{equation}

\noindent In   the large  $N$ limit   the  corresponding  density matrix
(\jtref{rho1}) reads

\begin{eqnarray}
\rho=   \frac{1}{{\cali{Z}}(\beta)}\exp\left(-\int   {d^{3}{\bf{k}}\over
(2\pi)^3 2\omega_k}\beta ({\bf k}){\tilde{\cal H}}({\bf k}) \right) \, ,
\jtleq{inicon1} 
\end{eqnarray}

\noindent   with   $\beta({\vect{k}})/2(2\pi)^{3}\omega_{k}$   being the
Lagrange multiplier to  be determined.  Using  Eq.(\jtref{jg3}), we find
that

\begin{eqnarray}
g({\bf      k})=  {V\over       (2\pi)^3}    \,      {\omega_k     \over
e^{\beta({\bf{k}})\omega_k}-1} \, , \jtleq{inicon2} 
\end{eqnarray}

\noindent where  $V$  denotes   the volume of   the   system.   Clearly,
expression (\jtref{inicon2}) can be interpreted as the density of energy
per mode.   The  fact that $\beta({\bf{k}})$  is  not constant indicates
that different modes have different `temperatures'.

The  Kadanoff-Baym equations  coincide  in   this  case with  those   in
(\jtref{em7}) provided  $m_{r}(T) \rightarrow {\cali{M}}$.  The boundary
condition  can  be    worked    out simply   using   the    prescription
(\jtref{CMS11}). This gives

\begin{equation}
G_{+-}(k) = e^{-\beta({\bf{k}})k_{0}}G_{-+}(k).  \jtleq{CMS2} 
\end{equation}

\noindent The fundamental solution of the Kadanoff-Baym equation is

\begin{eqnarray}
&&G_{\alpha  \beta}(k)= {(\sigma_{3})_{\alpha  \beta} \over  k^2 - {\cal
M}^2 +   i\epsilon_{\alpha \beta}} -  2\pi  i \delta ( k^2  -{\cal M}^2)
f_{\alpha     \beta}(k)\,  ,     \nonumber    \\    &&   \nonumber    \\
&&\mbox{\hspace{3cm}}   \epsilon_{\alpha      \beta}       =    \epsilon
(\sigma_{3})_{\alpha \beta}\,.  \jtlea{re2} 
\end{eqnarray}

\noindent Let    us  mention one more   point.   The  boundary condition
(\jtref{CMS2}) is  not  by  itself  sufficient to  determine  $f_{\alpha
\beta}$'s. (This fact is often overlooked  even in the equilibrium QFT.)
It is actually necessary to substitute this condition with an additional
condition, namely the condition of  causality.  The causality condition,
i.e.  vanishing of the commutator $[\Phi(x),  \Phi(y)]$ for $(x-y)^{2} <
0$, importantly restricts  the  class of possible $f_{\alpha  \beta}$'s.
To    see  this, let us    look at  the  Pauli-Jordan  function $\langle
\,[\Phi(x), \Phi(y)]\rangle$.  The latter is the homogeneous solution of
the Kadanoff-Baym equation with the initial-time value data: $\langle \,
[\Phi(x), \Phi(y)]\rangle|_{x_{0}=y_{0}}  =0$ (i.e. causality condition)
and $\partial_{0}\langle   \, [\Phi(x), \Phi(y)]\rangle|_{x_{0}=y_{0}} =
-\delta^{3}({\bf{x}}-{\bf{y}})$.  Thus    the Pauli-Jordan function   is
uniquely determined and its Fourier transform reads 

\begin{equation}
\mbox{F.T.}(\langle   [\Phi(x),\Phi(y)]     \rangle      )  =     -i2\pi
\delta(k^{2}-{\cali{M}}^{2})\; \varepsilon(k_{0}).  \jtleq{ft1} 
\end{equation}

\noindent Relation (\jtref{ft1}) immediately implies that 

\begin{eqnarray}
&&f_{+-}(k) = \theta(-k_{0}) + {\tilde{f}}(k) \nonumber \\ &&f_{-+}(k) =
\theta(k_{0}) + {\tilde{f}}(k)\,, \jtlea{f1} 
\end{eqnarray}

\noindent with ${\tilde{f}}$  being,   so far  arbitrary, and  for  both
$f_{+-}$ and $f_{-+}$ identical,   function. The ${\tilde{f}}$ is   then
fixed via the generalised  KMS condition (\jtref{CMS2}).  Similarly, the
causality   condition  specifies   that    $G_{++}(k)  -  G_{--}(k)    =
\mbox{PP}\{\, 1/(k^{2} - {\cali{M}}^{2})\}$ (the symbol `PP' denotes the
principal part).  By inspection of  the definition of $G_{\alpha \beta}$
one may easily realise that 

\begin{eqnarray}
&G_{++} +  G_{--}   =   G_{+-} + G_{-+}&\,,\nonumber   \\  &G_{+-}(k)  =
-(G_{+-}(-k))^{*}&\,, \nonumber \\ &G_{++}(k) = -(G_{--}(-k))^{*}&\,. 
\end{eqnarray}

\noindent  From   these  relations  follows  that  $f_{++}   =  f_{--} =
{\tilde{f}}$   and ${\tilde{f}}(k)    =  ({\tilde{f}}(-k))^{*}$.     The
${\tilde{f}}$ is the same as in (\jtref{f1}). 

Since the divergences in $G_{\alpha \beta}$  come from the first term in
(\jtref{re2}) (i.e.  from   the particular solution),  we can  shift the
corresponding (zero  temperature)  poles at  $d=4$ to  the bare mass. In
this case we can write 

\begin{equation}
{\cal M}^2 \equiv m^{2}_{r}+\delta m^2 \, , \jtleq{re33} 
\end{equation}

\noindent where  $m_{r}$ is  the  renormalised mass   in the vacuum  and
$\delta   m$   is  the mass    shift due   to  an  interaction  with the
non-equilibrium medium.  Inserting the  `$++$' components of (\ref{re2})
into (\ref{re3}), we obtain 

\begin{eqnarray}
{\cal M}^2=m^2_0+  {1\over   2}\lambda_0  \left[{{\cal M}^{{d-4}}  \over
(4\pi)^{d\over   2}} \Gamma\left[  {1-d/    2}   \right] \, +    N({\cal
M}^2)\right] \, , \jtleq{re4} 
\end{eqnarray}

\noindent where 

\begin{equation}
N({\cal M}^2)= \int {d^dk\over (2\pi)^d} 2\pi \delta ( k^2 -{\cal M}^2 )
f_{++}(k) \, . \jtleq{re5} 
\end{equation}

\noindent In an equilibrium  system $f_{++}$ would be  the Bose-Einstein
distribution.     Note  that because    (\jtref{re5})  corresponds  to a
homogeneous solution of the  Kadanoff-Baym  equation at $|x-y|=0$ it  is
automatically finite.  Thus for a translationally invariant medium (both
equilibrium and non-equilibrium) $f_{++}$ must act as a regulator in the
UV region.

Let    us  consider  the expression     for   the expectation  value  of
$\Theta^{\mu\nu}$ given in (\ref{me2}). It is a matter of a few lines to
show that 

\begin{eqnarray}
\langle  \Theta^{\mu\nu}\rangle_{\rm    ren} &=&iN    \int    {d^dk\over
(2\pi)^d}k^{\mu}k^{\nu}\left[G_{++}(k)-G(k)\right] \nonumber\\   &-&   i
{N\over      4}g^{\mu\nu}  \delta  m^2    \int    {d^dk\over   (2\pi)^d}
\left[G_{++}(k)+G(k) \right] \, , \jtleq{precal1} 
\end{eqnarray}

\noindent  with  $G$  being the  $T=0$   causal  Green function.   Using
(\jtref{re6}),  (\jtref{re3}) and  (\jtref{re2})  one may directly check
that  Eq.(\jtref{precal1}) does not   contain UV singularities when  the
limit  $d\to 4$ is  taken.  This verifies  our introductory remark, that
(\jtref{ppp10})   is finite even in   non-equilibrium  context. From the
generalised   KMS condition  (\jtref{CMS2})   and  from (\jtref{re2}) we
obtain that $f_{++}$ is

\begin{eqnarray}
f_{++}(k)={1\over e^{\beta({\bf{k}})\omega_k}-1} \, .  \jtleq{precal2} 
\end{eqnarray}

\noindent So far the results obtained were  completely general and valid
for any translationally invariant system.  Let  us now consider a system
in     which      $g({\bf     k})=    {V     \omega_k\over     (2\pi)^3}
\exp\left(-{\omega_k/\sigma}\right)$.  As we  shall  see, this condition
corresponds to  systems  where the  lowest frequency  modes  depart from
strict equilibrium, whilst     the   high energy ones  obey     standard
Bose-Einstein    statistics.   This  behaviour    is   typical of   many
off-equilibrium systems,   eg.   ionised atmosphere ,   laser stimulated
plasma, hot fusion  \jtcite{JD}, etc.  The  $\sigma$ is usually referred
to  as the  ionisation half-width  of  the  energetical  spectrum.  From
(\ref{inicon2})  we can determine $\beta({\bf k})$  as a function of the
physical parameter $\sigma$:

\begin{eqnarray}
\beta({\bf                                                          k})=
{1\over\sigma}+{1\over\omega_k}\sum_{n=1}^{\infty}{(-1)^{n+1}\over    n}
e^{-n\omega_k/\sigma} \,.  \jtleq{precal2a1} 
\end{eqnarray}

\begin{figure}[h]
\vspace{4mm} \epsfxsize=8.5cm \centerline{\epsffile{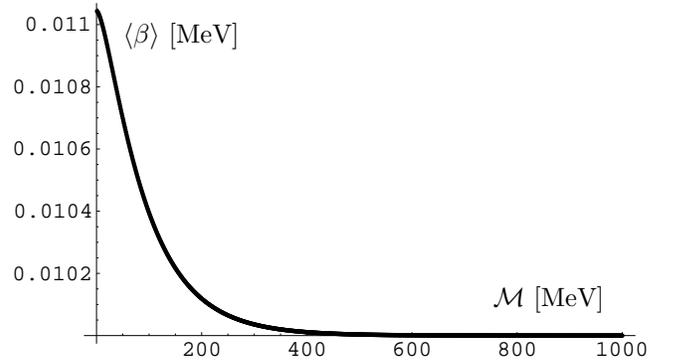}} 
\caption{\em  A plot  of  $\langle  \beta\rangle$ vs.  ${\cali{M}}$   at
$\sigma = 100$ \mbox{MeV}.} 
\label{fig2}
\begin{picture}(20,7)
\put(35,145){$\langle  \beta \rangle $ [MeV] } \put(175,45){${\cali{M}}$
[MeV]} 
\end{picture}
\end{figure}

\noindent   To   proceed,   some  remarks    on  the   interpretation of
$\beta({\bf{k}})$   are   necessary.   Firstly,   Eq.(\jtref{precal2a1})
implies   that  for a sufficiently    large $\omega_{k}$ ($\omega_{k}\gg
\sigma$) the  function $\beta({\bf{k}})$ is  approximately constant, and
equals $1/\sigma$.    Thus  at high  energies  the distribution $f_{++}$
approaches  the  Bose-Einstein distribution with the  global temperature
$T\approx \sigma$.  In  other words, only the  soft modes were sensitive
to   processes which    created   the non-equilibrium  situation.    The
interpretation  of  $\sigma $  as  an  equilibrium temperature, however,
fails whenever $\sigma  \approx \omega_{k}$.  Instead  of $\sigma $  one
may alternatively work with the  expectation value of $\beta({\bf{k}})$,
i.e. 

\begin{eqnarray}
\langle\beta\rangle   &=&  {\int    d^3{\bf    k}\,   \beta({\bf  k})e^{
-{\omega_{k}/\sigma}}\over     \int d^3{\bf k} \,e^{-{\omega_k/\sigma}}}
\nonumber\\    &=&   {1\over    \sigma}    +  {    {\sum}_{n=1}^{\infty}
{(-1)^{n+1}\over    n(n+1)}K_1({(n+1){\cali{M}}/\sigma})\over {\cali{M}}
K_2({{\cali{M}}/\sigma})} \, , \jtleq{precal2a} 
\end{eqnarray}

\noindent where $K_{n}$ is the Bessel  function of imaginary argument of
order $n$.   Because  the system  is  for  the significant part   of the
energetical  spectrum in thermal  equilibrium, $1/\langle \beta \rangle$
approximates the    corresponding    equilibrium  temperature   to  high
precision.  An interesting  feature of (\jtref{precal2a})  is that it is
quite insensitive to the  actual  value of ${\cali{M}}$.  Dependence  of
$\langle  \beta \rangle$ on  ${\cali{M}}$ for a  sample  value $\sigma =
100$MeV is depicted in FIG.\ref{fig2},

\vspace{3mm}

\noindent  An important observation is  that the asymptotic behaviour of
$\langle     \beta      \rangle$    at  $\sigma\to\infty$    goes   like
$\langle\beta\rangle\approx    a/\sigma$,        where      $a=1+{1\over
2}{\sum}_{n=1}^{\infty}  {(-1)^{n+1}\over   n(n+1)^2}=   \mbox{ln}2    +
\pi^{2}/24 \approx 1.1$.

Using (\jtref{ppp10}) and (\jtref{precal1}) we  get for the pressure per
particle

\begin{equation}
P  =  {1\over    2\pi^2}\left[   {\cali{M}}^2 \sigma^2  K_2({{\cali{M}}/
\sigma})        +            {\delta       m^2\over   4}{\cali{M}}\sigma
K_1({{\cali{M}}/\sigma})\right]- {\cali{P}}_0 \, , \jtlea{precal2b} 
\end{equation}

\noindent where 

\begin{eqnarray}
{\cali{P}}_0  &=&{1\over 384\pi^2}  (\delta   m^2)m_{r}^2\left(2+{\delta
m^2\over       m_{r}^2}\right)    \nonumber\\      &+&           {1\over
64\pi^2}m_{r}^4\left(1+{\delta           m^2\over        m_{r}^2}\right)
\ln\left(1+{\delta m^2\over m_{r}^2}\right) \, .  \jtleq{precal2c} 
\end{eqnarray}

\noindent  Note that ${\cali{P}}_{0}$  comes from the UV divergent parts
of   (\jtref{precal1}).    Whilst the  separate    contributions  are UV
divergent, they  cancel  between themselves  leaving  behind the  finite
${\cali{P}}_{0}$.  As we already  emphasised, the divergences  come from
the  particular solutions of  the Kadanoff-Baym  equations.  Because the
former  do  not directly reflect the  boundary  conditions, the  form of
${\cali{P}}_{0}$ must  be identical  for  any translationally  invariant
medium.  The non-trivial  information about the non-equilibrium pressure
is then encoded in terms in the brace $[\ldots]$ in (\jtref{precal2b}). 

For a  sufficiently  large  $\sigma$ the  leading behaviour  of  $\delta
m^{2}$  may be  easily evaluated.  To  do this,  let  us first  assemble
(\jtref{re6})  and (\jtref{re33})-(\jtref{re5}) together.  This gives us
the (renormalised) gap equation 

\begin{eqnarray}
\delta      m^{2}    &=&         \frac{1}{2          \lambda_{0}}\left\{
\frac{\Gamma[1-\frac{d}{2}]}{(4\pi)^{\frac{d}{2}}}\left({\cali{M}}^{d-4}
- m_{r}^{d-4}\right) + N({\cali{M}}^{2})\right\}\nonumber \\ 
&=& \frac{ \lambda_{r}}{2}\left\{   {\tilde{\Sigma}}(m_{r}^{2},   \delta
m^{2})                 +                        \frac{1}{2\pi^{2}}\sigma
{\cali{M}}K_{1}[{\cali{M}}/\sigma]\right\}, \jtleq{precal2e} 
\end{eqnarray}

\noindent with 

\begin{displaymath}
\frac{1}{2}   {\tilde{\Sigma}}(\ldots) =   \frac{1}{32 \pi^{2}}  \left\{
(m_{r}^{2}+      \delta      m^{2})\mbox{ln}\left(    1+    \frac{\delta
m^{2}}{m_{r}^{2}}\right) - \delta m^{2}_{r} \right\}. 
\end{displaymath}

\noindent Setting $x= \delta m^{2}/ m^{2}_{r}$ and $s  = \sigma / m_{r}$
we obtain the following transcendental equation for $x$

\begin{eqnarray}
\lambda_{r}^{-1}  &=&  \frac{1}{32  \pi^{2}}  \left\{ \frac{1}{x} \left[
(1+x) \mbox{ln}(1+x) - x \right. \right.\nonumber \\ && \nonumber \\ &&+
\left. \left.   8  s \sqrt{1+x}\;K_{1}[\sqrt{1+x}/s]\;  \right]\right\}.
\jtleq{trans1} 
\end{eqnarray}

\begin{figure}[h]
\vspace{4mm} \epsfxsize=8cm \centerline{\epsffile{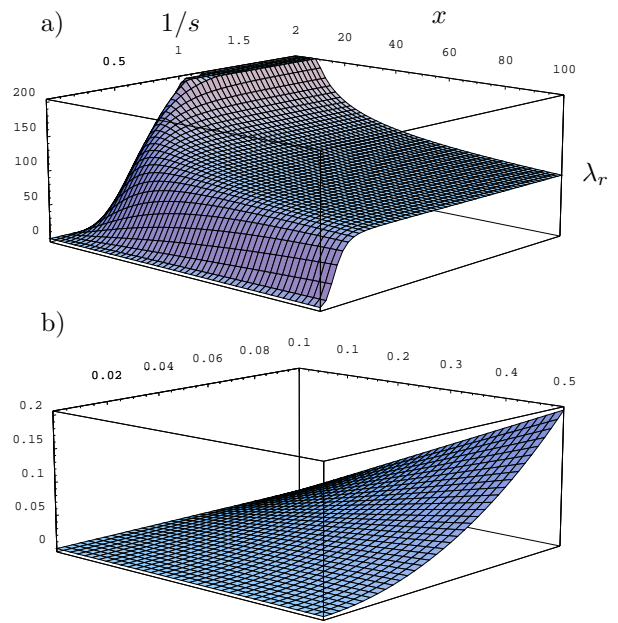}} 
\caption{\em A plot of the Eq.(\jtref{trans1}): a) the general shape, b)
a small $x$ behaviour.} 
\label{fig3}
\setlength{\unitlength}{1mm} 
\begin{picture}(20,7)
\put(6,95){a)}        \put(78,75){$\lambda_{r}$}        \put(58,96){$x$}
\put(22,95){$1/s$} \put(6,55){b)} 
\end{picture}
\end{figure}

\noindent   The   graphical representation of   Eq.(\jtref{trans1})   is
depicted in  FIG.\ref{fig3}.  From this one may  read off that for large
$x$  ($x > 50$)  there exists  a critical value  of  $\lambda_{r}$ above
which the gap  equation has no solution.   (The plateau is actually bent
downward with a very gentle  slope.)  FIG.\ref{fig3}b clearly shows that
if $\lambda_{r} \ll 1/s   < 1$ then $x  \ll  1$.  Using   the asymptotic
behaviour of $K_{1}[z]$ for $z \rightarrow 0$ ($K_{1} \sim (z)^{-1}$) we
arrive at more  precise estimate of $\lambda_{r}$  for which  $x \ll 1$;
namely $\lambda_{r}\approx  1/s^{2}   = m_{r}^{2}/  \sigma^{2}$.    This
estimate is very helpful for the asymptotic expansion of $\delta m^{2}$.
For a sufficiently high $\sigma$ we may write

\begin{eqnarray}
\delta     m^{2}  =    \frac{\lambda_{r}    \sigma^{2}}{4   \pi^{2}}   +
{\cali{O}}({\cali{M}}\mbox{ln}({\cali{M}}/\sigma)).  \jtleq{bb5} 
\end{eqnarray}

\noindent  Inserting this result into  (\jtref{precal2b})  and using the
series representation of  both   $K_2$ and $K_1$, keeping   the  leading
terms, we get

\begin{eqnarray}
P(\sigma)    &=&     \frac{\sigma^{4}}{\pi^{2}}    -  \frac{\sigma^{2}\,
{\cali{M}}^{2}}{2\pi^{2}}          +         \frac{\lambda_{r}}{8}\left(
\frac{\sigma^{2}{\mathcal{M}}^{2}}{64  \pi^{4}}  - \frac{\sigma^{4} 3}{4
\pi^{4}} \right)  \nonumber \\  && \nonumber \\  &+& {\mathcal{O}}\left(
{\mathcal{M}}^{2}{\mbox{ln}}({\mathcal{M}}/\sigma);      \lambda_{r}^{2}
\right).  \jtleq{bb6} 
\end{eqnarray}

\vspace{3mm}

\begin{figure}[h]
\vspace{4mm} \epsfxsize=8.5cm \centerline{\epsffile{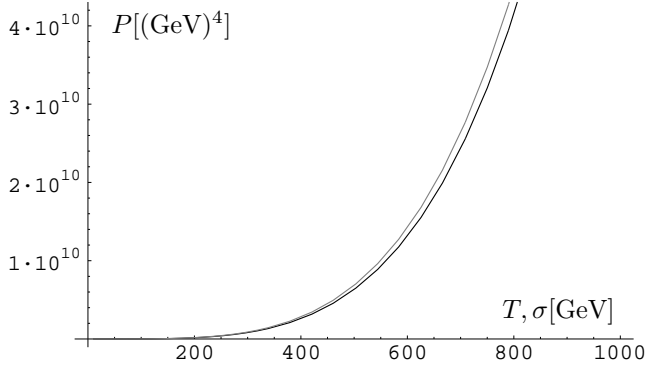}} 
\vspace{3mm} 
\caption{\em A plot  of pressure  as a function   of $T$,  $\sigma$  for
$m_{r}=  100\mbox{MeV}$.   The   gray line  corresponds   to equilibrium
pressure, the black line corresponds to pressure (\jtref{bb6}).} 
\label{fig5}
\setlength{\unitlength}{1mm} 
\begin{picture}(20,7)
\put(63,28){$T,      \sigma          \mbox{[GeV]}$}       \put(11,66){$P
[(\mbox{GeV})^{4}]$} 
\end{picture}
\end{figure}

It  is interesting to note  that $1/\pi^{2} \approx \pi^{2}/97$ which is
almost  the  Stefan-Boltzmann constant.  This   once more vindicates our
interpretation of $\sigma $ as a ``temperature''. A plot of the pressure
as a function of $\sigma $ is  depicted in FIG.\ref{fig5}.  It is due to
the low  frequence modes contribution to  the pressure that $P(\sigma) <
P(T) $.  This is a direct result  of our choice of $g({\bf{k}})$, namely
that $\sigma$ cannot be interpreted as temperature for the low frequence
modes  (i.e. $\omega_{k} < \sigma$).  The  smaller $\sigma$  is the less
important   contribution from  non-equilibrium  soft  modes  and so  the
smaller difference between both pressures. 

The  result  (\jtref{bb6}) can be  alternatively  rewritten in  terms of
 $\langle  \beta  \rangle$  .     Using,  for instance,   the   Pad\'{e}
 approximation \jtcite{bak} for $\langle \beta \rangle$, we arrive at

\begin{eqnarray}
P(\langle \beta  \rangle )  &=& 0.0681122\,{\langle \beta  \rangle^{-4}}
-0.0415368\,{\langle \beta \rangle^{-2}}{\cali{M}}^{2}\nonumber   \\ &+&
\lambda_{r}\,\left(      -0.000647\,{\langle    \beta   \rangle^{-4}}  +
0.0000164\,{\langle          \beta      \rangle^{-2}}\,{\mathcal{M}}^{2}
\right)\nonumber             \\         &+&          {\mathcal{O}}\left(
{\mathcal{M}}^{2}{\mbox{ln}}({\mathcal{M}}\langle     \beta    \rangle);
\lambda_{r}^{2} \right). 
\end{eqnarray}

\noindent The coefficient $0.0681122 \approx \pi^{2}/145$ is $1.6$ times
smaller than  the required value  for the Stefan-Boltzmann  constant, so
the parameter $\langle \beta  \rangle$ is a slightly worse approximation
of the equilibrium temperature than  $\sigma$.  In practice, however, it
is a matter of taste and/or a particular context  which of the above two
descriptions is more useful.

\begin{figure}[h]
\vspace{4mm} \epsfxsize=8.5cm \centerline{\epsffile{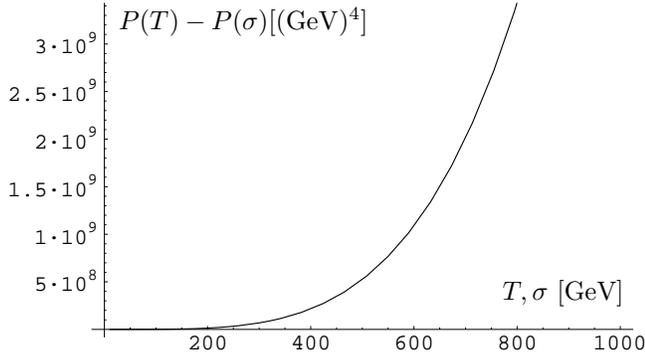}} 
\vspace{3mm} 
\caption{\em A  plot   showing   the  difference of   equilibrium    and
non-equilibrium pressures for $m_{r}=100\mbox{MeV}$.} 
\label{fig55}
\setlength{\unitlength}{1mm} 
\begin{picture}(20,7)
\put(63,25){$T,                   \sigma             $            [GeV]}
\put(12,61){$P(T)-P(\sigma)[(\mbox{GeV})^{4}]$} 
\end{picture}
\end{figure}

\subsection{Off-equilibrium II}

\noindent As was noted above, it is the  specific form of the constraint
$g({\bf{k}})$ which prescribes  the behaviour of $\beta({\bf{k}})$.  Let
us  now   turn our  attention to  systems  which depart  `slightly' from
equilibrium, i.e. when $g({\bf{k}})$ in (\jtref{ppp9}) deviates a little
from  the equilibrium  density of energy  per  mode.   In this  case the
constraint (\jtref{ppp9}) reads

\begin{equation}
g({\bf{k}})      =                 \frac{V}{(2               \pi)^{3}}\,
\frac{\omega_{k}}{e^{\beta_{0}\omega_{k}}-1}  + \delta g({\bf{k}}); \;\;
\delta g({\bf{k}}) \ll g({\bf{k}}), \jtlea{cond1} 
\end{equation}

\noindent with  $\beta_{0}=1/T_{0}$ being an  inverse of the equilibrium
temperature. As a special example of (\jtref{cond1}) we choose 

\begin{equation}
g({\bf{k}})          =                 \frac{V}{(2           \pi)^{3}}\,
\frac{\omega_{k}}{e^{\beta_{0}\omega_{k}}\alpha^{-1}({\bf{k}})-1};\;\;\;
\alpha({\bf{k}}, \beta_{0}) \approx 1.  \jtlea{cond2} 
\end{equation}

\noindent The inverse   mode ``temperature'' $\beta({\bf{k}})$ is   then
$\beta_{0}-          \mbox{ln}(\alpha({\bf{k}}))/\omega_{k}$.         So
$\mbox{ln}(\alpha)$ measures (in units of $\omega_{k}$) the deviation of
the mode temperature from the equilibrium  one.  The particular value of
$\alpha({\bf{k}}, \beta_{0})$  depends on the  actual  way in  which the
system is  prepared.  To  avoid unnecessary technical  complications, we
select  $\alpha ({\bf{k}},\beta_{0})$  to  be a  momentum-space constant
(generalisation is, however,  straightforward).  This choice  represents
the change in the  mode temperature which  is now inversely proportional
to $\omega_{k}$; the deviation is  bigger for soft  modes and is rapidly
suppressed for higher    modes.  Obviously, $T_{0}$ becomes the   global
temperature if  $\omega_{k} \gg \mbox{ln}(\alpha)$.  The generalised KMS
conditions      (\jtref{CMS2})   together  with  solutions (\jtref{re2})
determine $f_{++}$ as

\begin{eqnarray}
f_{++}(k)    ={\alpha  \over   {e^{\beta_0\omega_k}-\alpha}}    \,     .
\jtleq{precal3} 
\end{eqnarray}

\noindent Eq.(\jtref{precal3})  is   a reminiscent of   the, so  called,
J\"uttner   distribution\footnote{It   should  be   mentioned  that this
similarity is rather superficial.  The J\"{u}tner distribution describes
systems which are in thermal but not chemical equilibrium. (As we do not
have a chemical potential, chemical equilibrium is ill defined concept.)
The fugacity then parametrises  the deviation from chemical equilibrium.
}with   fugacity   $\alpha$  \jtcite{bairetal,strickland}.   Now,  using
(\jtref{ppp10}) and (\jtref{precal1})   we  get  for  the pressure   per
particle

\begin{eqnarray}
P    +   {\cal  P}_0  &=&    {1\over   2\pi^2}\left[  {{\cali{M}}^2\over
\beta_0^2}\sum_{n=1}^{\infty}       {\alpha^{n}\over     n^2}      K_2(n
{\cali{M}}\beta_0) \right.  \nonumber \\ &&+\left. {{\cali{M}}\delta m^2
\over    4\beta_0}\sum_{n=1}^{\infty}      {\alpha^{n}\over    n}  K_1(n
{\cali{M}}\beta_0)\right] \, , \jtleq{precal4} 
\end{eqnarray}

\noindent where $\delta m^2$ satisfies the gap equation 

\begin{eqnarray}
\delta     m^2  &=&{\lambda_r\over 2}\left(  {\tilde{\Sigma}}(m^{2}_{r},
\delta  m^{2}) +   {1\over    2\pi^2}\int_0^{\infty}{k^2dk\over\omega_k}
{\alpha \over  {e^{\beta_0\omega_k}-\alpha}}\right) \, .\nonumber  \\ &&
\jtleq{precal5} 
\end{eqnarray}

\noindent If we  set,  as  before,  $x=\delta m^{2}/m_{r}^{2}$ and   $s=
1/\beta_{0}m_{r}$ we get the transcendental equation for $x$

\begin{eqnarray}
\lambda^{-1}_{r}&=&  \frac{1}{32\pi^{2} x}\left\{ (1+x)\mbox{ln}(1+x) -x
\right.  \nonumber \\  &+&  \left.  8(1+x)\; \int^{\infty}_{1}\,   dz \,
\sqrt{z^{2}-1}  \, \frac{\alpha}{e^{z\,\sqrt{x+1}/s}- \alpha}  \right\}.
\jtleq{precal54} 
\end{eqnarray}

\begin{figure}[h]
\vspace{4mm} \epsfxsize=8cm \centerline{\epsffile{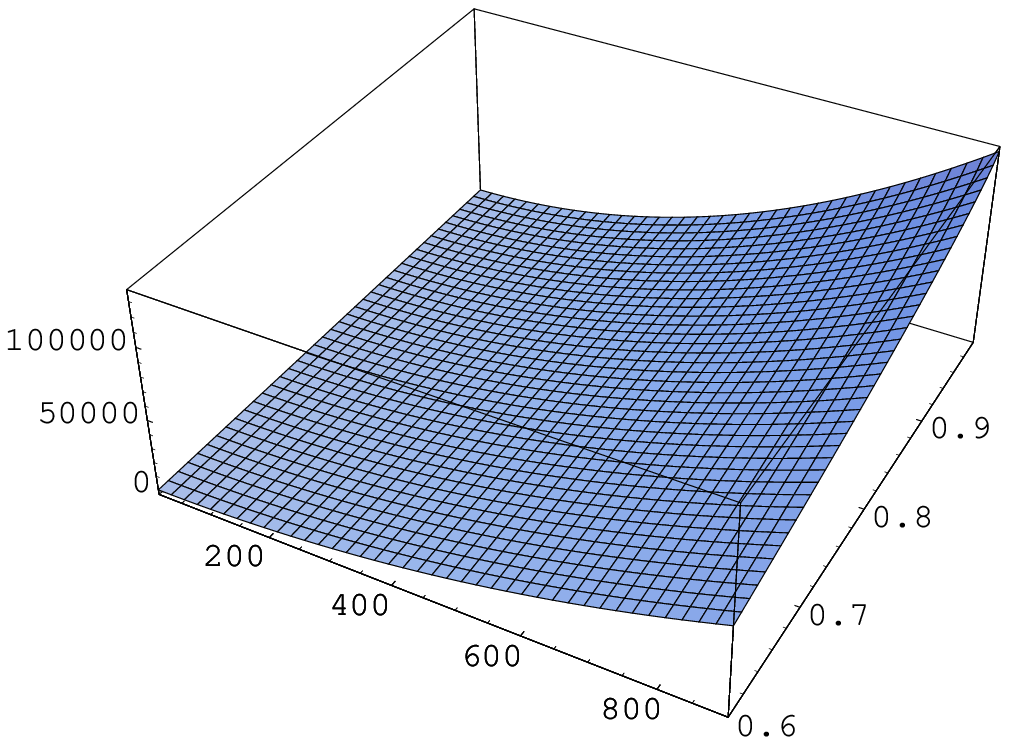}} 
\vspace{2mm} 
\caption{\em Behaviour of the pressure  (\jtref{precal4}) as a  function
of $\alpha$ and $T_{0}$ at $m_r=100 \mbox{MeV}$.} 
\label{fig76}
\setlength{\unitlength}{1mm} 
\begin{picture}(20,7)
\put(38.5,66){\small{$P     [(\mbox{GeV})^4]$}}    \put(74,30){$\alpha$}
\put(22,19.5){\small{$T_0 [\mbox{GeV}]$}} 
\end{picture}
\end{figure}

\noindent The  corresponding   numerical analysis  of (\jtref{precal54})
reveals that for $x \ll 1$ , $1/s \ll 1$.  So at $x \ll 1$, $\lambda_{r}
\approx   \delta m^{2}/T_{0}^{2}$.  This estimate   is important for the
asymptotic expansion of $\delta m^{2}$.   However, in order to carry out
the  asymptotic  expansion    of (\jtref{precal5})  (and    consequently
(\jtref{precal4}))   we need        to   cope first    with    the   sum
$\sum_{n=1}^{\infty}
\alpha^{n}K_{k}(n{\cali{M}}\beta_{0})/n^{k}\,\,\,(k=1,2)$  (also  called
Braden's function).  Expansion of $K_{k}(\ldots)$ yields a double series
which is very slowly convergent, and so it  does not allow one to easily
grasp the leading  behaviour in  $T_{0}$. In this  case it  is useful to
resume  (\jtref{precal4})-(\jtref{precal5})   by  virtue of  the  Mellin
summation technique  \jtcite{LW}.   (It is  well known \jtcite{LW,PJ,WH}
that  at equilibrium this resummation leads   to a rapid convergence for
high temperatures.) As a result, for a sufficiently large $T_{0}$ we get 

\begin{eqnarray}
\delta  m^2   &=&  \frac{\lambda_{r}\,T^2_{0}}{24}  -{\lambda_rT_0 {\cal
M}\over     4\pi}   \left[{1\over 2}(1-r^2)^{1/2}  \right.   \nonumber\\
&-&\left.                 r\left(1-\ln\left({{\cal               M}\over
2T_0}\right)\right)-(1-r^2)^{1/2}\arcsin    (r)\right] \nonumber\\   &+&
{\cal O}(\ln T_0) \; , \jtleq{precal6} 
\end{eqnarray}

\noindent where we have  set $r= \mbox{ln}(\alpha)T_{0}/{\cali{M}}$. The
corresponding expansion of the pressure (\jtref{precal4}) reads

\begin{eqnarray}
P  &= & {\pi^2  T_0^4\over 90} +{{\cal M}\zeta  (3) \over \pi^2} T_0^3 r
-{T_0^2 {\cal  M}^2 \over 24}\left(  1-2 r^{2}\right)  - {{\cal M}^3 T_0
\over  4\pi^2}   \nonumber\\  &\times &  \left[-  {1  \over  3} \left(1-
r^{2}\right)^{\frac{3}{2}}+          {r\over           2}\left(1-{2\over
3}r^2\right)\left(1   -    2\ln\left({{\cal            M}\over         2
T_0}\right)\right)\right.   \nonumber\\   &-   &   {2\over    9}\left(1-
r^{2}\right)\bigg(-r^3+3 - \left. 3 \left(1-  r^{2}\right)^{\frac{1}{2}}
\bigg)\arcsin    (r)\right]   \nonumber\\ &+&  {(\delta   m^2)^2   \over
2\lambda_r}+ {\cal O}(\ln T_0) \, .  \jtleq{precal6a} 
\end{eqnarray}

\noindent  where $\zeta(3)\approx 1.202$. Note that  for $\alpha = 1$ we
regain the equilibrium expansion (\jtref{b9}). The corresponding plot of
$P$  as  a     function  of  $T_{0}$   and   $\alpha$ is    depicted  in
FIG.\ref{fig76}.

\noindent In    passing    it may be    mentioned    that the  expansion
 (\ref{precal6a}) is mathematically justifiable only for $\alpha \approx
 e^{\sqrt{\lambda_{r}/24}\, r} \approx 1$.

\vspace{3mm} 


\section{Summary and Conclusions}
\label{sec6}

\noindent
In   order  to  get a   workable   recipe for  non-equilibrium  pressure
calculations   we  have    combined  three   independent concepts:   the
Jaynes-Gibbs   principle  of  maximal   entropy,   the   Schwinger-Dyson
equations,  and   the hydrostatic pressure.   The  basic  steps   are as
follows. 

To find quantitative results for pressure one needs to know the explicit
form of  the Green's functions involved. These  may be find  if we solve
the Dyson-Schwinger  equations.  The  corresponding solutions are unique
provided  we specify the  density   matrix $\rho$ (the construction   of
$\rho$ is one of  the cornerstones of our approach,  and we tackled this
problem   using the Jaynes-Gibbs  principle  of  maximal entropy).  With
$\rho$  at  our disposal  we   showed how to  formulate  the generalised
Kubo-Martin-Schwinger (KMS) boundary conditions.

To show how the   outlined method works  we  have illustrated  the whole
procedure on an exactly  solvable model, namely $O(N)\; \Phi^{4}$ theory
in the large $N$ limit.   This model is  sufficiently simple yet complex
enough  to  serve as  an illustration  of   basic characteristics of the
presented method in contrast to other ones in use.  In order to find the
constraint  conditions we  have  considered two  gedanken experiments in
which  the system in question  was prepared in such  a way that only low
frequence modes  departed from the  strict  equilibrium behaviour.  Such
processes can   be found, for example, in   ionised atmosphere, in laser
stimulated   plasma  or in   hot  fusion \jtcite{JD}.    In both alluded
gedanken experiments we were  able to work  out the hydrostatic pressure
exactly.    Furthermore,  after identifying  a ``temperature'' parameter
(virtually temperature of  high modes) we  carried out the corresponding
high-temperature expansions.

As it  was  discussed, one  of  the main advantages  of the Jaynes-Gibbs
construction is that one  starts  with the (physical) constraints  (i.e.
parameters    which     are    really   controlled     and measured   in
experiments). These constraints   directly determine the  density matrix
and  hence   the  generalised KMS   conditions for   the Dyson-Schwinger
equations.  This contrasts with the  usual approaches where the  density
matrix is treated as the primary object.  In these cases it is necessary
to solve the von Neumann-Liouville equation.  This is usually circumvent
using  either  variational methods  \jtcite{EJY,FJ}  with several  trial
$\rho$'s or reformulating the problem in  terms of the quantum transport
equations  for Wigner's functions  \jtcite{Bal}.   It is, however,  well
known that  the inclusion of constraints into  transport equations  is a
very delicate  and rather complicated task (the  same  is basically true
about the variational methods) \jtcite{Bal,Nem}. 

\acknowledgements 
 

The authors are indebted to P.V.Landshoff  and D.A.Steer for reading the
manuscript and for helpful  discussion. One of  us  (EST) would like  to
thank to P.V.Landshoff,  J.C. Taylor, and  the HEP group  for their kind
hospitality in DAMTP.  This work was  partially supported by Fitzwilliam
College and CONACYT. 



\end{document}